\documentclass[5p,number,sort&compress]{elsarticle}                            %% Final print

\usepackage[utf8]{inputenc}
\usepackage[T1]{fontenc}
\usepackage{amsmath}
\usepackage{amssymb}
\usepackage{textcomp}
\usepackage[caption=false]{subfig}
\usepackage{booktabs}

\newcommand{\un}[1]{\,{\rm #1}}
\newcommand{\eps}{\varepsilon}
\newcommand{\dd}{\mathrm{d}}
\newcommand{\degree}{^\circ}
\DeclareMathOperator{\cov}{cov}
\DeclareMathOperator{\diag}{diag}

\usepackage{figurewidth}
\setfigurewidth{0.6}{1}

\journal{Astroparticle Physics}

\begin{document}

\begin{frontmatter}

\title{An Unfolding Method for X-ray Spectro-Polarimetry}

% This fixes the problem that all authors get a corresponding author asterisk
% Taken from http://tex.stackexchange.com/questions/116515/elsarticle-frontmatter-corresponding-author
\makeatletter
\def\@author#1{\g@addto@macro\elsauthors{\normalsize%
    \def\baselinestretch{1}%
    \upshape\authorsep#1\unskip\textsuperscript{%
      \ifx\@fnmark\@empty\else\unskip\sep\@fnmark\let\sep=,\fi
      \ifx\@corref\@empty\else\unskip\sep\@corref\let\sep=,\fi
      }%
    \def\authorsep{\unskip,\space}%
    \global\let\@fnmark\@empty
    \global\let\@corref\@empty  %% Added
    \global\let\sep\@empty}%
    \@eadauthor={#1}
}
\makeatother

\author{F.~Kislat\corref{cor}}
\ead{fkislat@physics.wustl.edu}
\author{M.~Beilicke}
\author{Q.~Guo}
\author{A.~Zajczyk}
\author{H.~Krawczynski}
\address{Washington University in St. Louis, Department of Physics and McDonnell Center for the Space Sciences, One Brookings Dr., CB 1105, St.\ Louis, MO 63130}

\cortext[cor]{Corresponding author}

\begin{abstract}
  X-ray polarimetry has great scientific potential and new experiments, such as X-Calibur, PoGOLite, XIPE, and GEMS, will not only be orders of magnitude more sensitive than previous missions, but also provide the capability to measure polarization over a wide energy range.
  However, the measured spectra depend on the collection area, detector responses, and, in case of balloon-borne experiments, the absorption of X-rays in the atmosphere, all of which are energy dependent.
  Combined with the typically steep source spectra, this leads to significant biases that need to be taken into account to correctly reconstruct energy-resolved polarization properties.
  In this paper, we present a method based on an iterative unfolding algorithm that makes it possible to simultaneously reconstruct the energy spectrum and the polarization properties as a function of true photon energy.
  We apply the method to a simulated X-Calibur data set and show that it is able to recover both the energy spectrum and the energy-dependent polarization fraction.
\end{abstract}


\end{frontmatter}

%%% Change when tagging a version:
%\noindent\textbf{Draft --- \today}
%\noindent\textbf{Version 1.2}

\section{Introduction}
X-ray polarimetry holds the promise to resolve the inner regions of compact systems like mass accreting black holes in X-ray binaries and X-ray bright neutron stars~\cite{krawczynski_h_2011}.
For example, spectropolarimetric observations of pulsars and pulsar wind nebulae can constrain the geometry and locale of particle acceleration in these sources.
Measurements of the polarization of X-rays from the Crab Nebula indicate an increase in the polarization fraction and a change in polarization angle in the energy range between a few keV and $100\un{keV}$ indicating that the~$\gamma$-ray emission must come from a small, highly ordered region, whereas X-rays are emitted from all morphological features of the pulsar wind nebula.
Spectropolarimetric observations can constrain the magnetic structure of jets in Gamma Ray Bursts and Active Galactic Nuclei.
Furthermore, X-ray polarimetry can be used to measure the masses and spins of black holes and the orientation of their inner accretion disk, as well as accretion disks and accretion disk coronae of Active Galactic Nuclei.
See Ref.~\cite{krawczynski_h_2011} and references therein for more details.

While this potential has long been appreciated, the OSO-8 satellite launched in 1978 has been the only mission with a dedicated X-ray polarimeter so far that measured X-ray polarization of an astrophysical source~\cite{weisskopf_m_c_1978}.
New technological developments enabled the design of compact wide-bandpass polarimeters with a large collection area such as the proposed satellites GEMS~\cite{hill_je_2012} and XIPE~\cite{soffitta_p_2013} or the balloon-borne hard X-ray polarimeters X-Calibur~\cite{guo_q_2013} and PoGOLite~\cite{pearce_m_2012}.
The collection areas and detection efficiencies of these experiments will allow spectropolarimetric observations with unprecedented sensitivity.
In this paper we study statistical methods that can be used to analyze the data of such experiments.

The measurement of the polarization fraction and direction of the abovementioned experiments make use of the photo-electric effect (GEMS, XIPE) or the Compton effect (PoGOLite, X-Calibur).

Photoelectrons are emitted preferentially parallel to the electric field of the electromagnetic wave associated with the photon.
The differential cross section of the photoelectric effect in the non-relativistic case can be approximated as~\cite{heitler_w_1936}:
\begin{equation} \label{eq:heitler}
  \frac{\dd\sigma}{\dd\Omega} = r_0^2\,Z^5\,\alpha^4 \left(\frac{m_ec^2}{E}\right)^{\frac{7}{2}} \frac{4\sqrt{2} \sin^2\theta \cos^2\phi}{(1 - \beta\cos\theta)^4},
\end{equation}
where $r_0$ is the classical electron radius, $Z$ the atomic number of the absorbing material, $\alpha$ the fine-structure constant, $m_e$ the electron rest mass, $E$ the photon energy, $\theta$ the angle between the incoming photon and the emitted photoelectron, $\beta$ its speed in units of $c$, and $\phi$ the azimuth angle of the emitted electron with respect to the polarization direction of the incident X-ray.

Photons scatter preferentially perpendicular to the electric field, as governed by the Klein-Nishina cross section (see e.\,g.\ Ref.~\cite{evans_r_1955}):
\begin{equation} \label{eq:klein-nishina}
  \frac{\dd\sigma}{\dd\Omega} = \frac{r_0^2}{2}\frac{k_1^2}{k_0^2} \left[\frac{k_0}{k_1} + \frac{k_1}{k_0} - 2\sin^2\theta\cos^2\eta\right],
\end{equation}
where $\eta$ is the angle between the electric vector of the incident photon and the scattering plane, $\mathbf{k}_0$ and $\mathbf{k}_1$ are the wave-vectors before and after scattering, and $\theta$ is the scattering angle.

The azimuthal scattering angle or emission direction of the photoelectron is, therefore, a proxy for the polarization angle.
In general, scattering polarimeters measure the azimuthal scattering angle of an X-ray photon and the energy of the scattered photon.
In some realizations an X-ray mirror is used to focus the beam onto the detector assembly.

Since typical source spectra exhibit steep power-laws, the energy resolution and energy-dependent detection efficiency have to be taken into account.
The most important effects that must be considered are:
\begin{itemize}
 \item the energy resolution of the detector;
 \item the energy lost in the scattering process, which is not measured by all experiments;
 \item the energy-dependent effective area of grazing incidence mirrors;
 \item absorption of photons in the atmosphere in case of balloon-borne instruments;
 \item the energy-dependent detection efficiency.
\end{itemize}

In this paper we describe an unfolding algorithm, and show that it can be used to determine flux and polarization fraction and direction as a function of photon energy with small biases.

In Section~\ref{sec:the_problem} we will define the problem.
In Section~\ref{sec:spectropolarimetry} we will introduce an unfolding method that can be used to reconstruct the photon spectrum of the source while preserving the energy-dependent azimuth distribution of events.
In Section~\ref{sec:example}, we apply the method to a set of simulated X-Calibur data.
Finally, in Section~\ref{sec:summary}, we summarize our findings and present our conclusions.

\section{Formulation of the problem} \label{sec:the_problem}
Typical X-ray polarimeters measure the energy of the scattered photon or the energy of the recoil electron.
This energy will differ from the energy of the incident photon by an unknown amount governed by underlying physical process (e.\,g.\ photoelectron emission or Compton scattering) and the finite energy resolution of the detectors.

\begin{figure}
  \centering
  \includegraphics[width=\figurewidth]{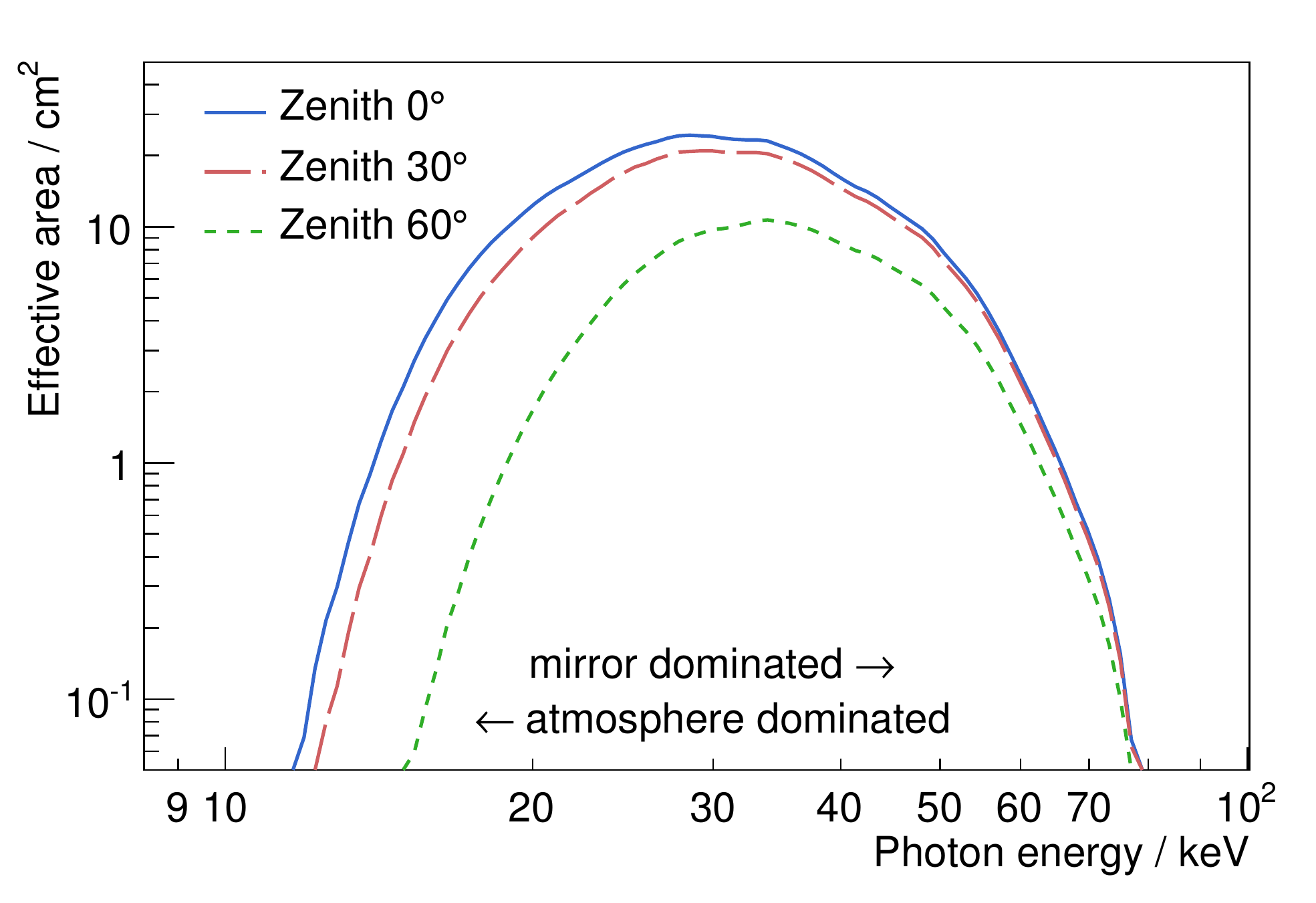}
  \caption{Effective collection area of a balloon-borne polarimeter flown at an altitude of ${\sim}45\un{km}$ (atmospheric depth of~$2.6\un{g\,cm^{-2}}$ at~$90\degree$ elevation), taking into account the absorption of X-rays in the residual atmosphere. At low energies the limiting factor is absorption in the atmosphere, while at high energies the mirror effective area limits the collection area.}
  \label{fig:effective_area}
\end{figure}

Grazing-incidence X-ray mirrors focus the X-rays from a source onto a detector and make it possible to combine large detection areas with rather small detectors, and thus to achieve excellent signal to background ratios.
In addition, they change the polarization properties of the incident photons by~${<}1\%$~\cite{sanchez_almeida_j_1993,katsuta_j_2009}.
However, their effective areas depend strongly on the energies of incident photons.
In case of a balloon-borne experiment, the atmosphere above the detector absorbs low-energy photons.
For illustration, Fig.~\ref{fig:effective_area} shows the effective collection area of the InFOC$\mu$S mirror~\cite{tueller_j_2005}, folded with the energy-dependent transmissivity of the atmosphere (from~\cite{photon_absorption}) for a flight altitude of~$45\un{km}$ ($2.6\un{g\,cm^{-2}}$ at~$90\degree$ elevation).

Together, the strong energy dependence of the effective collection area, the finite energy resolution, and the steeply falling source spectra, lead to a significant distortion of the measured energy spectra, which needs to be taken into account in a spectropolarimetric analysis.

Mathematically, these effects can be described by a convolution of a true spectrum with a detector response function which includes effects of the mirror and the atmosphere.
If the spectrum is measured in $n_e$ discrete energy bins, the measurement can be expressed as a vector of event counts $N^e_i\ (i=1, \ldots, n_e)$ and the detector response is represented by the response matrix~$\mathbf{R}$, which relates the true number of photons $N^c_j\ (j=1, \ldots, n_c)$ in the $j$-th energy bin to the observed counts in the $i$-the bin:
\begin{equation}\label{eq:fold_spectrum}
  N^e_i = \sum_{j=1}^{n_c} R_{ij} \, N^c_j.
\end{equation}
Each entry in the response matrix corresponds to the probability to measure an energy in bin~$i$ given a true energy in bin~$j$:
\begin{equation}
  R_{ij} = P(E^e_i|E^c_j).
\end{equation}
More generally, what was refered to as ``true spectrum'' so far is a set of ``\emph{causes}'', $\Phi^c_j$, with a set of properties -- one of them being the energy of the incident photon --, which lead to a set of ``\emph{effects}'', $\Phi^e_i$, with a set of properties that can be measured in the experimental setup (the ``measured spectrum'') -- one of them for example the energy of the scattered photon.
Therefore, instead of the ``measured spectrum'' we will from now on refer to the measured \emph{effects}.
The response matrix $R_{ij}$ then describes the probability that cause $j$ will lead to effect $i$.

In general, neither the binning of nor the number of properties associated with $\boldsymbol\Phi^e$ and $\boldsymbol\Phi^c$ need to be the same.
Thus, it is straight-forward to generalize equation~\eqref{eq:fold_spectrum} to include information about the azimuthal scattering angle~$\phi$ in cause bins $k=1, \ldots, n_\phi^c$ and effect bins $\ell=1, \ldots, n_\phi^e$ respectively.
Then, each \emph{cause} represents a combination of true energy and scattering angle, $\Phi^c_j = (E^c_p, \phi^c_k)$, $j = 1, \ldots, n_E^c\,n_\phi^c$, and correspondingly each \emph{effect} represents a combination of measured energy and scattering angle.
The vectors~$\boldsymbol{N^c}$ and~$\boldsymbol{N^e}$ now represent event numbers in bins of these more general, higher-dimensional causes~$\Phi_j^c$ and effects~$\Phi_i^e$.

Additional observables (i.\,e.\ parameters of the \emph{effects}) or parameters of the \emph{causes} (i.\,e.\ output parameters of the unfolding) can be added in the same way.
For instance, when analysing the X-Calibur data in Section~\ref{sec:example}, we add the coordinate of the observed photon along the optical axis as an additional input parameter in order to improve the energy resolution.

The response matrix is normalized such that for a given \emph{cause} bin $j$ the sum over all \emph{effect} bins is the detection efficiency $\eps_j$ for photons in the $j$-th \emph{cause} bin:
\begin{equation}\label{eq:response_properties}
  \sum_i R_{ij} = \eps_j,
\end{equation}
with $0 \leq \eps_j \leq 1$.

In general, measurements will contain a significant fraction of background events caused by cosmic rays and al\-be\-do photons~\cite{guo_q_2013}, despite active and passive shielding of the detector.
If the background distributions of the input variables are known -- ideally, they should be measured during flight in special off-source data taking runs --, the background can simply be subtracted from the input distributions.
This is possible because the unfolding method does not consider individual events but only takes distributions of measured quantities as input.
Therefore, in the remainder of this paper, we will neglect any background and only consider signal events.

According to the Klein-Nishina cross section, Eq.~\eqref{eq:klein-nishina}, photons scatter preferentially perpendicular to their electric field vector; and according to Eq.~\eqref{eq:heitler} photoelectrons are emitted preferentially parallel to the polarization vector.
This results in a sinusoidal modulation of the azimuthal scattering distribution with a~$180\degree$ period with minima at the direction of the polarization plane, and a relative amplitude proportional to the degree of polarization.
A common method to determine polarization parameters is to fit a sine function to the azimuthal distribution.
The modulation amplitude
\begin{equation}\label{eq:modulation-amplitude}
  \mu = \frac{C_\mathrm{max}-C_\mathrm{min}}{C_\mathrm{max}+C_\mathrm{min}}
\end{equation}
is therefore a measure of the polarization fraction.
$C_\mathrm{max}$ and $C_\mathrm{min}$ are the maximum and minimum of the sinusoidal modulation of the azimuthal distribution.
The performance of a polarimeter can be characterized by the modulation obtained when observing a~$100\%$ polarized source,~$\mu_{100}$, called the \emph{modulation factor}.

\section{Energy-resolved polarimetry}\label{sec:spectropolarimetry}

\subsection{Unfolding the energy spectrum}\label{sub:unfolding}
Typically, the response matrix~$\mathbf{R}$ is poorly conditioned and simply inverting it to obtain an estimate of~$\boldsymbol{N^c}$ from a measured spectrum~$\boldsymbol{N^e}$ leads to unnatural fluctuations of the result (see e.\,g.\ Ref.~\cite{hansen_pc_1998}).
Instead, we use an iterative unfolding technique~\cite{dagostini_1995}, which avoids this undesired effect.
In the following we summarize this method in the application to spectropolarimetric data.

Starting from a prior distribution $P^{(r)}(\Phi^c_j)$ in the $r$-th iteration, the posterior conditional probability $P^{(r)}(\Phi^c_j|\Phi^e_i)$ is calculated using Bayes' Theorem:
\begin{equation}\label{eq:unfoldingmatrix}
  P^{(r)}(\Phi^c_j|\Phi^e_i) = 
    \frac{P(\Phi^e_i|\Phi^c_j) \, P^{(r)}(\Phi^c_j)}{\sum_k P(\Phi^e_i|\Phi^c_k) \, P^{(r)}(\Phi^c_k)}.
\end{equation}
Then, an estimate of the true spectrum is obtained:
\begin{equation}
  \hat{N}^{c(r)}_j = \frac{1}{\eps_j}\sum_i N_i^e P^{(r)}(\Phi^c_j|\Phi^e_i),
\end{equation}
where $\eps_j$ is the efficiency in \emph{cause} bin $j$ as defined in Eq.~\eqref{eq:response_properties}.
At the end of each iteration $P^{(r)}(\Phi^c_j)$ is replaced by
\begin{equation}
  P^{(r+1)}(\Phi^c_j) = \frac{\hat{N}^{c(r)}_j}{\sum_p \hat{N}^{c(r)}_p}.
\end{equation}
This procedure is repeated for a number of iterations, $n_\mathrm{iter}$, which has to be determined in advance, typically using Monte Carlo simulation studies (see Section~\ref{sub:n_iter}).

Unlike suggested in Ref.~\cite{dagostini_1995}, $\hat{N}_j^{c(r)}$ was not smoothed between iterations in the studies presented in this paper.
Smoothing may result in a ``nicer'' looking result.
However, it can potentially introduce additional systematic uncertainties if the results depend on the smoothing kernel being used.

The initial prior $P^{(0)}(\Phi^c_j)$ is chosen either based on prior knowledge of the source or on the observed spectrum.
One can either start from a flat azimuthal distribution or from an azimuthal distribution derived from the observed data.
Since the unfolding result might depend on the choice of prior, the influence of this choice should be studied by repeating the unfolding with a range of ``reasonable'' priors.

The result of this unfolding is a two-dimensional distribution of the best estimate of event numbers as a function of incident photon energy and scattering angle.
A projection of this distribution onto the energy axis yields the source spectrum.
Slices along the azimuth axis can be fitted with a sine function in order to obtain polarization fraction and angle for individual energy bins.

\subsection{Determination of the response matrix} \label{sub:response_matrix}
In general, the response of the detector, $\mathbf{R}$, can be determined from Monte Carlo simulations.
These simulations have to accurately describe all parts of the experiment:
\begin{itemize}
 \item the absorption of X-ray photons in the atmosphere;
 \item the effective collection area of the mirror;
 \item the physical processes in the Compton scattering medium;
 \item the response of the photon detectors;
 \item and the resolution of the readout electronics.
\end{itemize}

For most astrophysical applications, it is reasonable to simulate a powerlaw spectrum of incident photons.
For each simulated photon that triggers a detector, one then determines the observed electronic signal at the readout electronics.
In this way, the distribution of measured observables is determined for each true cause bin, and the response matrix is found by dividing all of these distributions by the number of incident photons in the respective cause bin.

The absorption of photons in the atmosphere depends on the zenith angle of the observation and the altitude of the balloon, both of which vary with time.
Therefore, an observation of a source has to be considered as a sequence of observations $I = 1,\ldots,n_\mathrm{obs}$ of duration $T_I$, described by individual response matrices $\mathbf{R}_I$. Assuming that the source spectrum $N^c_j$ is constant during the entire observation time $T = \sum_I T_I$, the detector response can be described by an average response matrix:
\begin{equation} \label{eq:time_avg_response}
  N^e_i = \sum_{j=1}^{n_cn_\phi^c} \left(\frac{1}{T} \sum_{I=1}^{n_\mathrm{obs}} T_I R_{ij,I}\right) N^c_j.
\end{equation}
Since the individual observations only differ by the absorption in the atmosphere, the response matrices will only differ by a factor in each column:
\begin{equation} \label{eq:transmissivity_response}
  \mathbf{R}_I = \mathbf{R}_0\diag(\boldsymbol{t}_I).
\end{equation}
Therefore, absorption in the atmosphere as a function of photon energy in time interval $I$, $\boldsymbol{t}_I$, can be separated from the other effects listed above.
The vector $\boldsymbol{t}_I$ contains the probability that a photon in the $j$-th true energy bin will not be absorbed in the atmosphere:
\begin{equation} \label{eq:transmissivity}
  t_{I,j} = \exp(-d\,\mu(E_j)/\rho),
\end{equation}
where $d$ is the slant depth in $\mathrm{g\,cm^{-2}}$ and $\mu(E_j)/\rho$ is the mass attenuation coefficient in air within energy bin $E_j$.
Thus, it is sufficient to simulate the detector response excluding atmospheric absorption once ($\mathbf{R}_0$) and then apply the absorption effects for each time interval.

\subsection{Termination of the unfolding iteration} \label{sub:n_iter}
The optimal number of unfolding iterations has to be determined before attempting to unfold the experimental data in order to avoid potential biases.
Generally, the unfolded spectrum becomes a better representation of the true spectrum with increasing number of iterations.
However, at the same time fluctuations of the data are amplified leading to the same problem introduced by simply inverting the response matrix.
This can be understood since in some way the unfolding matrix from Eq.~\eqref{eq:unfoldingmatrix} is an ``inverse'' of the response matrix.
Since the response matrix smears fluctuations due to the limited detector response, the unfolding matrix will have the opposite effect, amplifying fluctuations.
In addition each new prior is based on the unfolding result from the previous iteration, whose fluctuations in this way feed back into the unfolding loop.
The iteration should thus be terminated when a compromise between a good representation of the true spectrum and small artificial fluctuations is reached.

One way to achieve this is to fold a hypothetical input spectrum and azimuthal distribution with the response matrix and then sample the same number of events as in the experimental data set from this distribution.
These ``\emph{measured}'' data are then unfolded to find the number of iterations that yields the best agreement between the result and the known \emph{true} input.
This should be repeated for various different inputs to find an optimum independent of the true spectrum and polarization fraction or angle.

The disadvantage of this method is that the ideal number of iterations may depend considerably on whether the true spectrum is smooth or contains features such as emission or absorption lines.
The choice of number of iterations may therefore lead to a significant systematic bias.

A better approach is to use the convergence of the unfolding process itself to determine when to stop the iteration.
After each iteration, the unfolded spectrum is folded with the response matrix, $\tilde{N}^{e(r)}_i = \sum_j R_{ij}\hat{N}^{c(r)}_j$, and compared to the measured data, $N_i^e$. 
A convergence criterion can then be defined using the change in~$\chi^2$ between~$\tilde{N}^{e(r)}_i$ and the measured data between two iterations~$r$ and~\mbox{$r+1$}~\cite{ulrich_2004}:
\begin{equation}\label{eq:delta_chi2}
  \Delta \tilde\chi^2(r, r+1) = \chi^2(\tilde{\boldsymbol{N}}^{e(r)}, \boldsymbol{N}^e) - \chi^2(\tilde{\boldsymbol{N}}^{e(r+1)}, \boldsymbol{N}^e).
\end{equation}
This quantity decreases monotonically during the iteration process.
However, after a large number of iterations as $\Delta\chi^2(r, r+1) \to 0$ the unfolding would introduce the same large-scale fluctuations as simply inverting~$R_{i,j}$.
To avoid this, the iteration is terminated once $\Delta\chi^2(r, r+1)$ falls below a certain value~$\Delta\chi^2_\mathrm{term}$.
The value of this limit can be determined beforehand using a simple Monte Carlo simulation similar to what would be used to determine a fixed maximum number of iterations as described above.
The best value of~$\Delta\chi^2_\mathrm{term}$ generally depends less on the shape of the true spectrum than the number of iterations.

\subsection{Statistical uncertainties of the unfolded spectrum} \label{sub:bootstrap}
Statistical uncertainties of the measured data are propagated through the unfolding process using a bootstrap method: the number of events within each bin of the input data is varied randomly $r=1,\ldots,n$ times according to a Poisson distribution with the measured value as the mean, and the unfolding is repeated.
The statistical error in bin $j$ is then determined by comparing each unfolding result $\hat{N}^{c(r)}_j$ to the average $\langle \hat{N}^c \rangle_j$:
\begin{equation}\label{eq:sigma}
  (\sigma^c_j)^2 = \frac{1}{n-1} \sum_{r=1}^n \bigl(\hat{N}^{c(r)}_j - \langle \hat{N}^c\rangle_j\bigr)^2.
\end{equation}
Likewise, bin-to-bin correlations are obtained:
\begin{equation}
  \cov(i,j) = \frac{1}{n} \sum_{r=1}^n \bigl(\hat{N}^{c(r)}_i - \langle \hat{N}^c\rangle_i\bigr)\bigl(\hat{N}^{c(r)}_j - \langle\hat{N}^c\rangle_j\bigr).
\end{equation}

\subsection{Implementation}\label{sub:implementation}
For the studies in the next sections we made the assumption that the azimuth angle of the scattered photons can be determined perfectly.
This means that the response matrix is diagonal in azimuth, which makes it a sparse matrix, essentially reducing its size by one dimension.
The authors have, therefore, implemented the unfolding algorithm using sparse matrix routines from the \texttt{SuiteSparse} package~\cite{suitesparse}, which lead to a significant performance improvement, compared to an implementation using dense matrix algorithms.
Even when taking into account uncertainties in the azimuth distribution, the response matrix will very likely still be sparse, meaning sparse matrix algorithms are likely a good choice in any case.

\section{Application to simulated X-Calibur data} \label{sec:example}

In this section we will apply the method described before to a simulation of a $6$-hour observation of the Crab nebula with the polarimeter X-Calibur flown in the focal plane of the InFOC$\mu$S telescope.

\subsection{Simulation setup}\label{sec:example_setup}

\begin{figure}
  \centering
  \includegraphics[width=\thisfigurewidth{.6}{.9}]{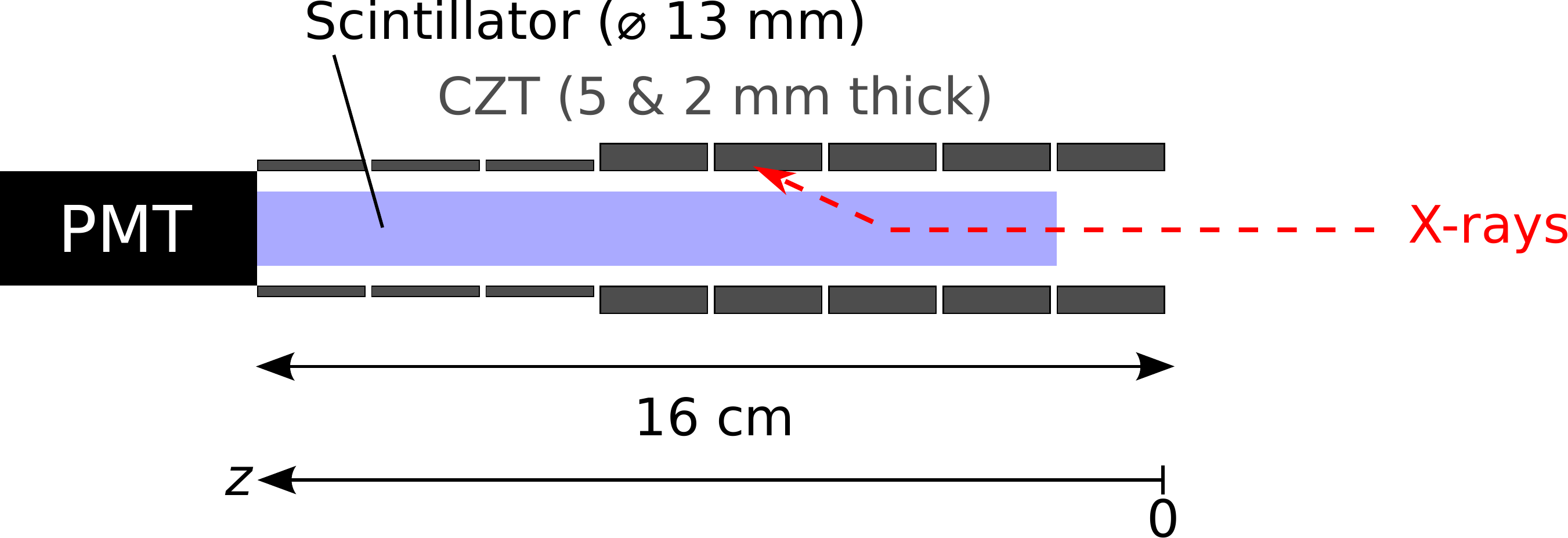}
  \caption{Principle of the X-Calibur polarimeter. A grazing incidence X-ray mirror focuses the X-rays onto the polarimeter. In this sketch the X-rays enter the detector from the right and scatter off a scintillator slab. The scattered X-rays are then detected by one of the CZT detectors that surround the scintillator. A coincidence requirement between the trigger from the photomultiplier tube at the end of the scintillator and events in the CZTs can be used to suppress backgrounds, but is not used in the analysis presented here.}
  \label{fig:xcalibur-principle}
\end{figure}

The detection principle of X-Calibur is illustrated in Fig.~\ref{fig:xcalibur-principle}.
Incoming X-rays Comp\-ton-scat\-ter in a~$14\un{cm}$ long, $13\un{mm}$ diameter scintillator rod, and the scattered photons are detected by 32 $2\un{mm}$ and $5\un{mm}$ thick Cadmium-Zinc-Telluride (CZT) detectors, each occupying an area of $20 \times 20\un{mm^2}$ divided into $8 \times 8$ pixels~\cite{guo_q_2013}.
In order to reduce systematic uncertainties, X-Calibur continuously rotates around the optical axis.
Observables are the azimuthal angle $\phi$ and energy $E$ of the scattered photon, and the coordinate $z$ along the optical axis at which the scattered photon was detected.

The $z$-coordinate of the detected photon is related to the scattering angle $\theta$ and the depth along the axis of the scatterer at which the Compton scattering occurred, which, however, cannot be measured in the experiment.
Therefore, $z$ carries important information on the loss of energy in the scatterer.
Including this variable in the analysis, therefore, provides valuable information helping to reconstruct the energy of the incident photon.

In order to test the unfolding method, the response of the X-Calibur polarimeter to incoming X-rays was simulated using Geant4~\cite{geant4,geant4_2}.
The response of the CZT detector was then simulated by propagating a charge proportional to the amount of energy deposited inside the detector through the electric field of the detector and integrating the induced charge at the electrodes.
Additionally, the electronic response was smeared with a Gaussian distribution with a resolution of~$10\%$.
The detector threshold was modeled as a hard cutoff at~$20\un{keV}$ measured energy.
The X-ray mirror was simulated by assigning a weight to each event proportional to the effective collection area at the incident photon energy.

Three data sets were generated.
The first (\texttt{DS1}) consisted of~$10^6$ horizontally polarized photons with an $E^{-1}$ spectrum in the range between~$10$ and~$85\un{keV}$ to be used to extract the response matrix.
It took into account the detector response and mirror effective area as described above but no atmospheric absorption.

\begin{figure}
  \centering
  \includegraphics[width=\figurewidth]{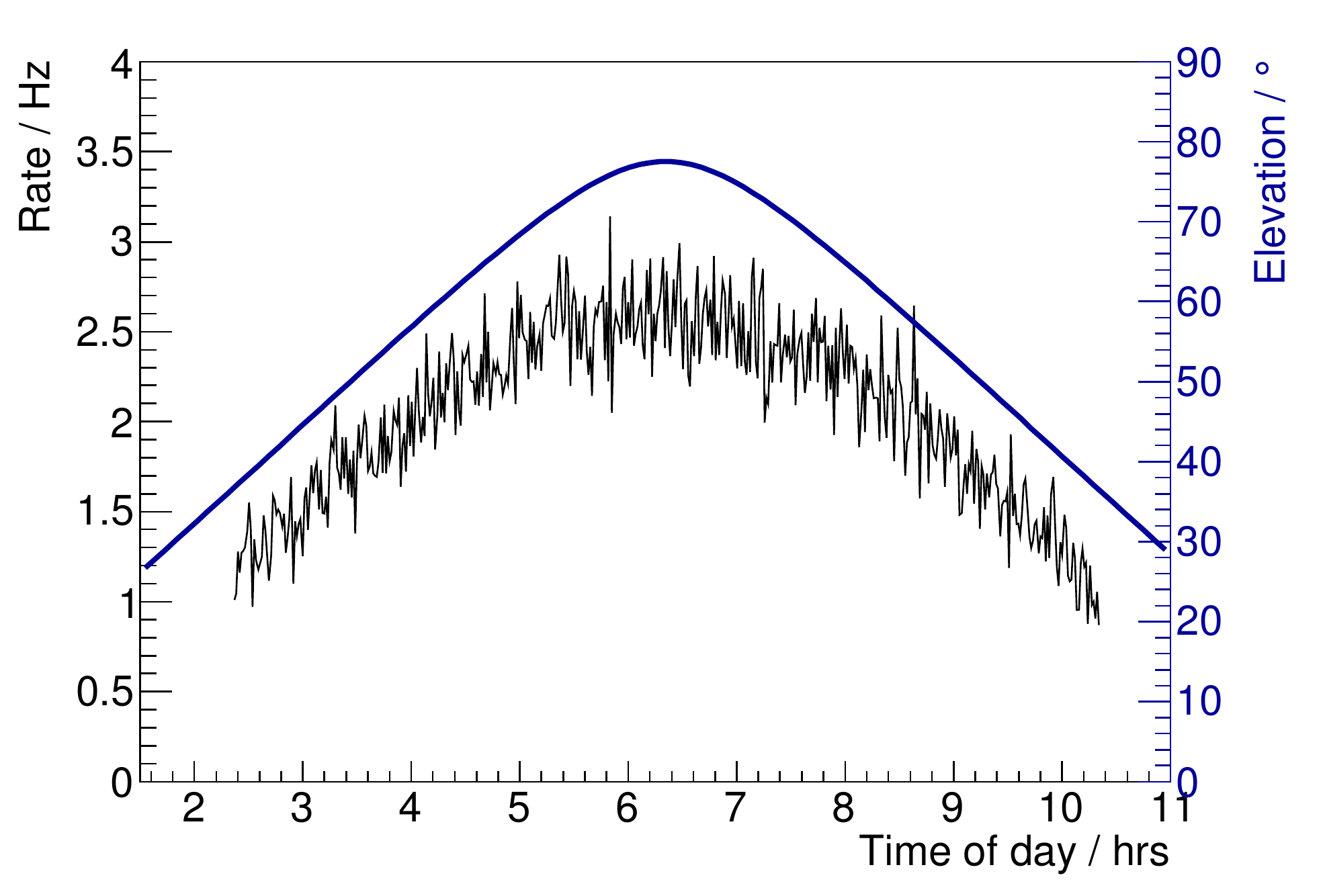}
  \caption{Simulated event rate in X-Calibur (thin line) and elevation of the Crab pulsar (bold line) as a function of time on 9/24/2014.}
  \label{fig:rate_vs_time}
\end{figure}

The second dataset (\texttt{DS2}) was used to test the unfolding.
It consisted of the same detector simulation but with a steeper incident spectrum, $E^{-2.15}$, as found in the Crab Nebula~\cite{tueller_j_2010}. 
Additionally, observation time dependent absorption of photons in the atmosphere (due to the change of source elevation) was introduced to simulate the observation of an astrophysical source (in our example the Crab Nebula as observed from Ft. Sumner, NM, USA) at an atmospheric overburden of~$2.6\un{g\,cm^{-2}}$.
The absorption as a function of energy was calculated from a spline interpolation of the NIST XCOM tables for air~\cite{photon_absorption}.
The polarization of the Crab Nebula has been measured at $2.6\un{keV}$, $5.2\un{keV}$~\cite{weisskopf_m_c_1978} and ${>}100\un{keV}$~\cite{dean_aj_2008,forot_m_2008}.
The polarization seems to increase from ${\sim}20\%$ at $2.6$ and $5.2\un{keV}$ to values of $46\%$~\cite{dean_aj_2008} or even ${>}77\%$~\cite{forot_m_2008} at ${>}100\un{keV}$.
We assume an energy dependent polarization fraction $p(E)$ of:
\begin{equation}\label{eq:truepolarization}
  p(E) = 0.65 - \frac{0.45}{\exp\left(\frac{E/\un{keV} - 40}{5}\right) + 1},
\end{equation}
which leads to $p \sim 20\%$ at $20\un{keV}$ and $p \sim 65\%$ above $70\un{keV}$ with a smooth transition around~$40\un{keV}$.
The polarization angle does not change in this example.
The number of events was chosen to match an $8$-hour observation of the Crab Nebula with X-Calibur assuming a flux of
\begin{equation}\label{eq:truespectrum}
  F(E) = 10.17 \left(\frac{E}{1\un{keV}}\right)^{-2.15} \mathrm{cm^{-2}\,s^{-1}\,keV^{-1}}
\end{equation}
as measured by the Swift Burst Alert Telescope~\cite{tueller_j_2010}.
Each event in this dataset was assigned a time of day, which progressed to match the expected rate.
The zenith angle of the source at this time was stored with each event.
The observation time dependence of the event rate that resulted from this zenith angle variation is shown in Fig.~\ref{fig:rate_vs_time}.
This dataset contained a total of~23,872 events.

\begin{figure}
  \centering
  \includegraphics[width=\figurewidth]{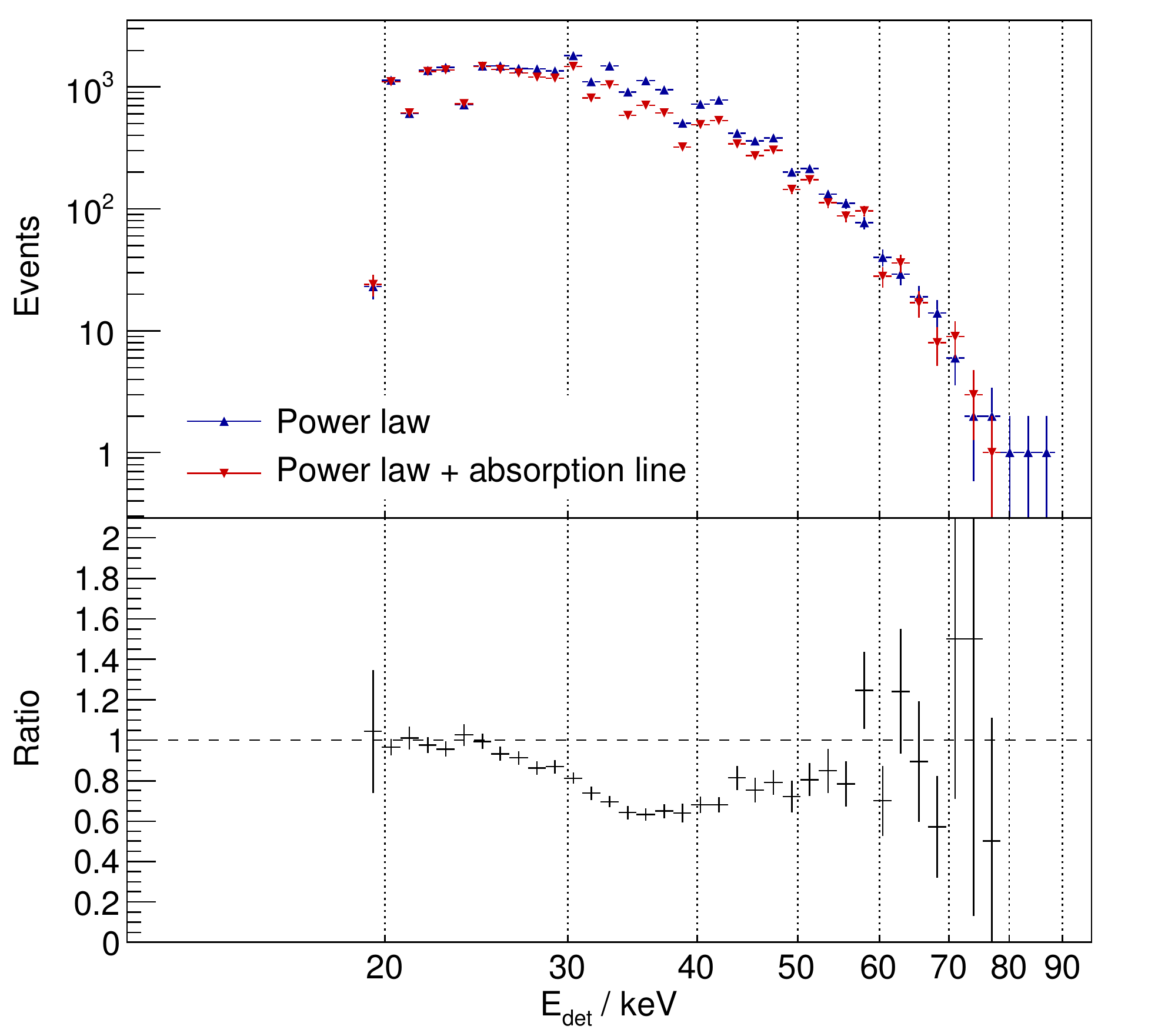}
  \caption{Spectrum of measured photon energies in data sets \texttt{DS2} (pure power-law) and \texttt{DS3} (power-law + absorption line). The bottom panel shows the ratio of the two spectra highlighting the effect of the absorption line. The large bin-to-bin variations at low energies are a binning effect: the discretization of energies in the detector is linear whereas the binning in these histograms was done on a logarithmic scale.}
  \label{fig:line-noline-comparison}
\end{figure}

A third dataset (\texttt{DS3}) was created from the second one to include an absorption line at an energy of~$E_\mathrm{line} = 40\un{keV}$ by discarding events with a probability of
\begin{equation}
 p_\mathrm{absorb}(E_\mathrm{phot}) = 0.5 \times \exp\left(-\frac{(E_\mathrm{phot} - E_\mathrm{line})^2}{2 \sigma_\mathrm{line}^2}\right)
\end{equation}
with a width of $\sigma_\mathrm{line} = 5\un{keV}$, which is comparable to the $10\%$~resolution of the simulated instrument.
Figure~\ref{fig:line-noline-comparison} shows the spectra of measured photon energies~$E_\mathrm{det}$ for datasets \texttt{DS2} and \texttt{DS3} as well as their ratio, which clearly shows the effect of the absorption line.

\subsection{Response matrix and modulation factor}\label{sub:example_response_matrix}
From data set \texttt{DS1} the response matrix~$\mathbf{R}_0$ was constructed as described in Section~\ref{sub:response_matrix}.
Using the tables in Ref.~\cite{photon_absorption} and the elevation angle of the Crab (see Fig.~\ref{fig:rate_vs_time}), the atmospheric absorption in each energy bin was calculated in $1\un{s}$ intervals.
These absorption coefficients were then applied to the response matrix~$\mathbf{R}_0$ using Eqns.~\eqref{eq:time_avg_response} and~\eqref{eq:transmissivity_response}.

In this example we consider the simplified case where the response is diagonal in azimuth, i.\,e. we assume $\phi_e = \phi_c$, and measured and ``true'' azimuth angles will be binned in the same way, $n_\phi^c = n_\phi^e =: n_\phi$.
The azimuth needs to be included in the response matrix in order to reconstruct the dependence of the azimuth distribution on the true energy in the unfolding.
Assuming $\phi_e = \phi_c$ is not a conceptual restriction and can be removed easily. 
It is merely a simplification used in the study of X-Calibur data in this paper because the precision at which the scattering angle is measured does not limit the reconstruction of the polarization angle.

\begin{figure}
  \centering
  \includegraphics[width=\thisfigurewidth{0.6}{0.8}]{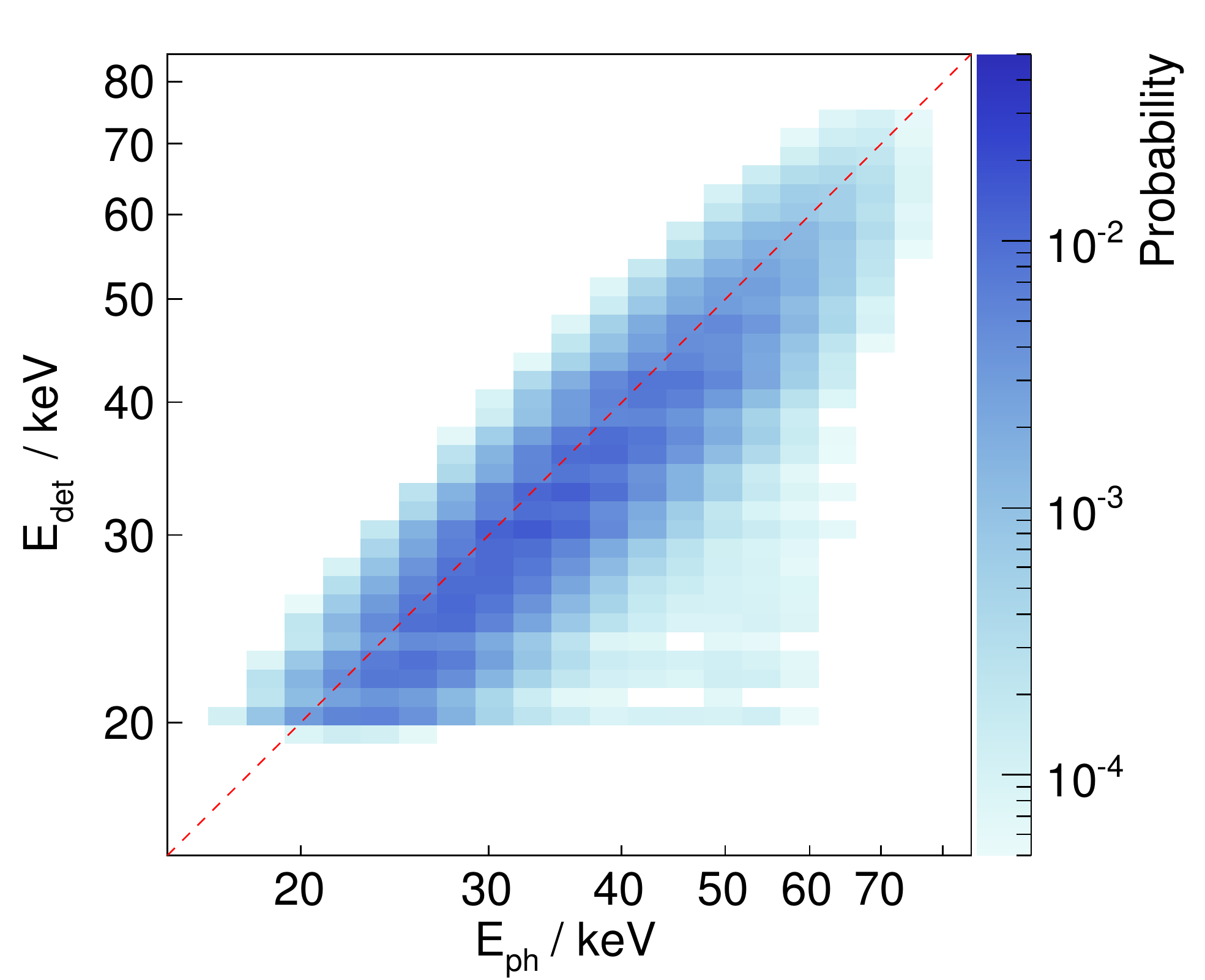}
  \caption{Projection of the response matrix onto the $E_\text{detected}$-vs.-$E_\text{photon}$ plane.}
  \label{fig:response_matrix}
\end{figure}

A representation of the energy response of X-Calibur is shown in Fig.~\ref{fig:response_matrix}.
On average, the detected energy is less than the incident photon energy as the photon transfers some of its energy to the Compton electron.
The distribution of measured energies shows a long tail towards low values.
This is due to the combined effect of the physics of Compton scattering, fluorescence photons escaping the CZTs, the depth-dependent charge collection efficiency of the CZT detectors, and occasional Compton scattering rather than photoelectric effect interactions in the CZT crystals.
At high energies, the efficiency is dominated by the small effective area of the mirror, while at low energies absorption in the atmosphere is the dominant effect.

Finally, the modulation factor was determined as a function of energy.
In each energy bin the azimuthal scattering angle distribution measured for a~$100\%$ polarized beam was fitted with a sine function,
\begin{equation}\label{eq:sine}
  f(\chi) = A \sin(n\chi - \phi) + B,
\end{equation}
with the free parameters: amplitude~$A$, periodicity~$2\pi/n$, phase~$\phi$, and offset~$B$.
The modulation factor is then
\begin{equation}
  \mu_{100} = A/B.
\end{equation}
No energy dependence of~$\mu_{100}$ was found and an average value of $\langle\mu_{100}\rangle = 0.498 \pm 0.003$ was determined.

\subsection{Number of unfolding iterations}
In order to determine the optimal number of iterations, the unfolding was applied to simulated data distributions.
A true energy spectrum and energy-dependent scattering angle distribution was folded with the response matrix determined above.
The result is the 3-dimensional distribution of detected energies, scattering angles and $z$ coordinates one would find in the absence of statistical errors.
This distribution was then scaled such that its integral corresponded to the total number of events in the test dataset to be unfolded (\texttt{DS2} and \texttt{DS3}).

The event number~$N$ in each bin was then randomized according to a normal distribution with mean~$N$ and standard deviation~$\sqrt{N}$.
This simulated dataset was then unfolded with the same prior that was going to be used to unfold the simulated data.
After each iteration
\begin{itemize}
 \item the statistical errors of the unfolded spectrum were calculated according to Eq.~\eqref{eq:sigma};
 \item $\Delta\tilde\chi^2$ was calculated according to Eq.~\eqref{eq:delta_chi2};
 \item and the $\chi^2$ between the unfolded and the true spectrum was calculated.
\end{itemize}
This process was repeated~$100$ times.

\begin{figure}
  \centering
  \includegraphics[width=\figurewidth]{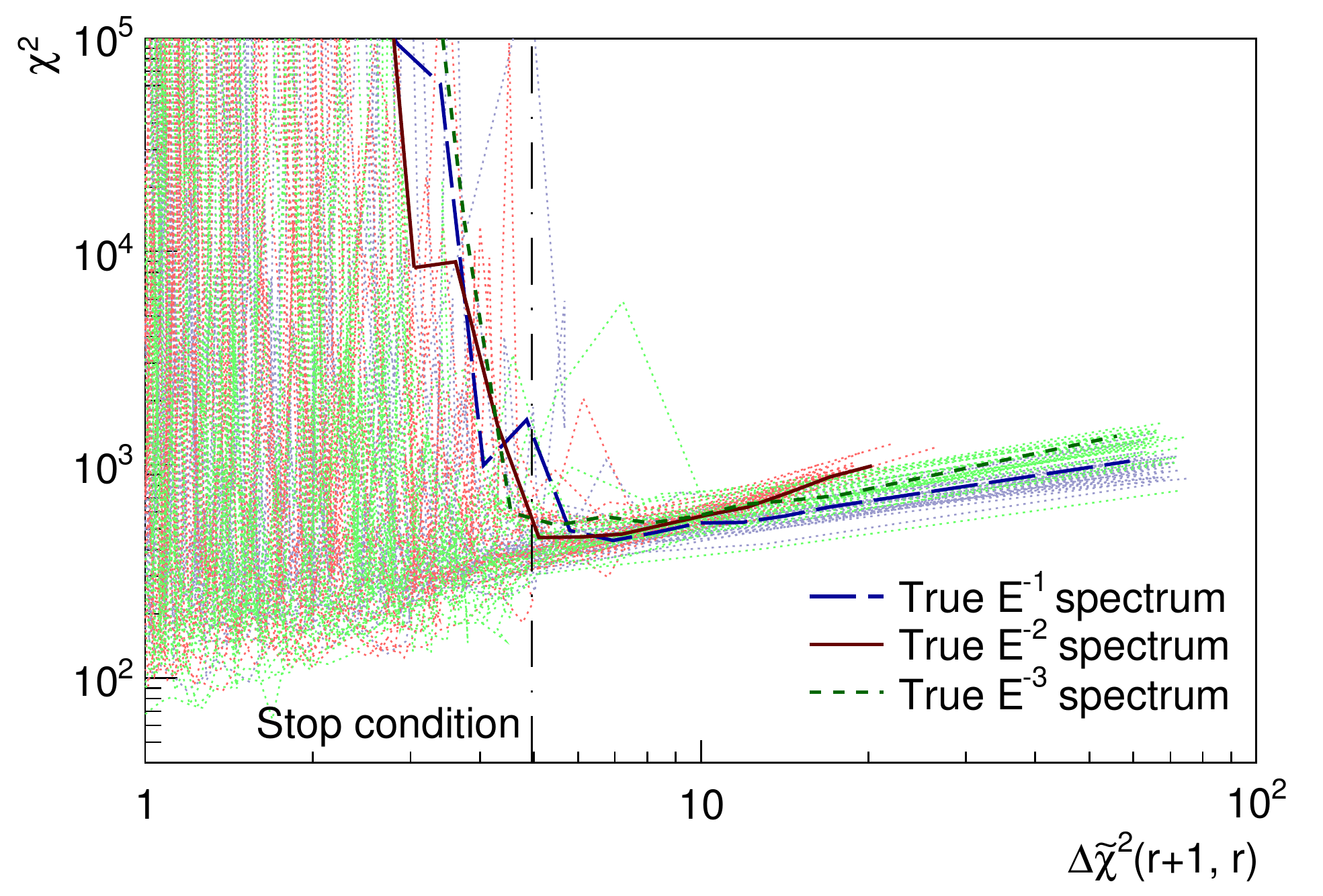}
  \caption{Difference between true and unfolded spectrum as a function of $\Delta\tilde\chi^2$, defined in Eq.~\eqref{eq:delta_chi2}, as described in Section~\ref{sub:n_iter}. The unfolding proceeds from right to left as $\Delta\tilde\chi^2$ decreases monotonically. At first, $\chi^2$ decreases until reaching a minimum after which it increases again due to numerical noise. The different colors represent true spectra with three different powerlaw indices. In all cases an energy-independent polarization fraction of~$0.5$ was assumed. The prior was in all cases an~$E^{-2}$ spectrum flat in azimuth. The dotted lines represent individual unfolded spectra. For each true spectrum 30 randomizations are shown. The thick lines are the average over all 100 variations per true spectrum. When applied to data the iteration will proceed until $\Delta\tilde\chi^2$ drops below the value indicated by the vertical dash-dotted line labeled ``Stop condition''.}
  \label{fig:niter}
\end{figure}

The result of this procedure for three different true spectra with an energy-independet polarization fraction of~$0.5$ is shown in Fig.~\ref{fig:niter}.
In all cases an~$E^{-2}$ prior with a flat azimuthal distribution was chosen.
Generally, the agreement between the unfolded and the true spectrum improves with increasing number of iterations, i.\,e.\ decreasing~$\Delta\tilde\chi^2$.
However, when exceeding a certain number of iterations, the $\chi^2$ starts to vary by many orders of magnitude.
At this point, the unfolding procedure starts to amplify the statistical fluctuations.
Generally, the unfolding iteration should be stopped before reaching this region.
More detailed analysis shows that the greatest improvement is reached within the first few iterations.

Two unfolding tests were performed, both based on the same input described in Sect.~\ref{sec:example_setup}.
First, the data were unfolded starting from an $E^{-2}$ prior spectrum, then with an $E^{-1}$ prior.
Assuming the true spectrum was unknown, the optimal number of iterations for three different true spectra each with polarization fractions of $0$, $0.5$, and $1$ was determined using the method described in this section.
Furthermore, spectra with an absorption line with a width of $5\un{keV}$ at $E=40\un{keV}$ were studied.
The decision on the number of iterations for the two examples was then reached based on curves similar to those in Fig.~\ref{fig:niter}.
We chose to iterate until $\Delta\tilde\chi^2 < 5$, which resulted in the number of iterations listed in Table~\ref{tab:iterations} for the four cases studied here.

\begin{table}
  \centering
  \caption{Number of unfolding iterations resulting from the stopping condition $\Delta\tilde\chi^2 < 5$ for the two priors and the two true spectra.}
  \begin{tabular}{lrr}
    \toprule
    \textbf{Prior index:}     & $\mathbf{-1}$ & $\mathbf{-2}$ \\
    \midrule
    \textbf{Power-law }       & 3             & 5             \\
    \textbf{Power-law + line} & 6             & 6             \\
    \bottomrule
  \end{tabular}
  \label{tab:iterations}
\end{table}

\subsection{Unfolding results}
To test the unfolding method, the simulated data sets \texttt{DS2} and \texttt{DS3} were first unfolded starting from a prior that followed an~$E^{-2}$ energy spectrum and was flat in azimuth for all energies.
In a second test, an~$E^{-1}$~prior, again flat in azimuth, was used.
Figure~\ref{fig:unfolding_result} shows the spectra of measured energies, the true spectra and the energy spectra obtained as a result of the unfolding.

\begin{figure*}[p]
  \centering
  \subfloat[Power-law starting from~$E^{-2}$ prior.]{
    \includegraphics[width=0.45\textwidth]{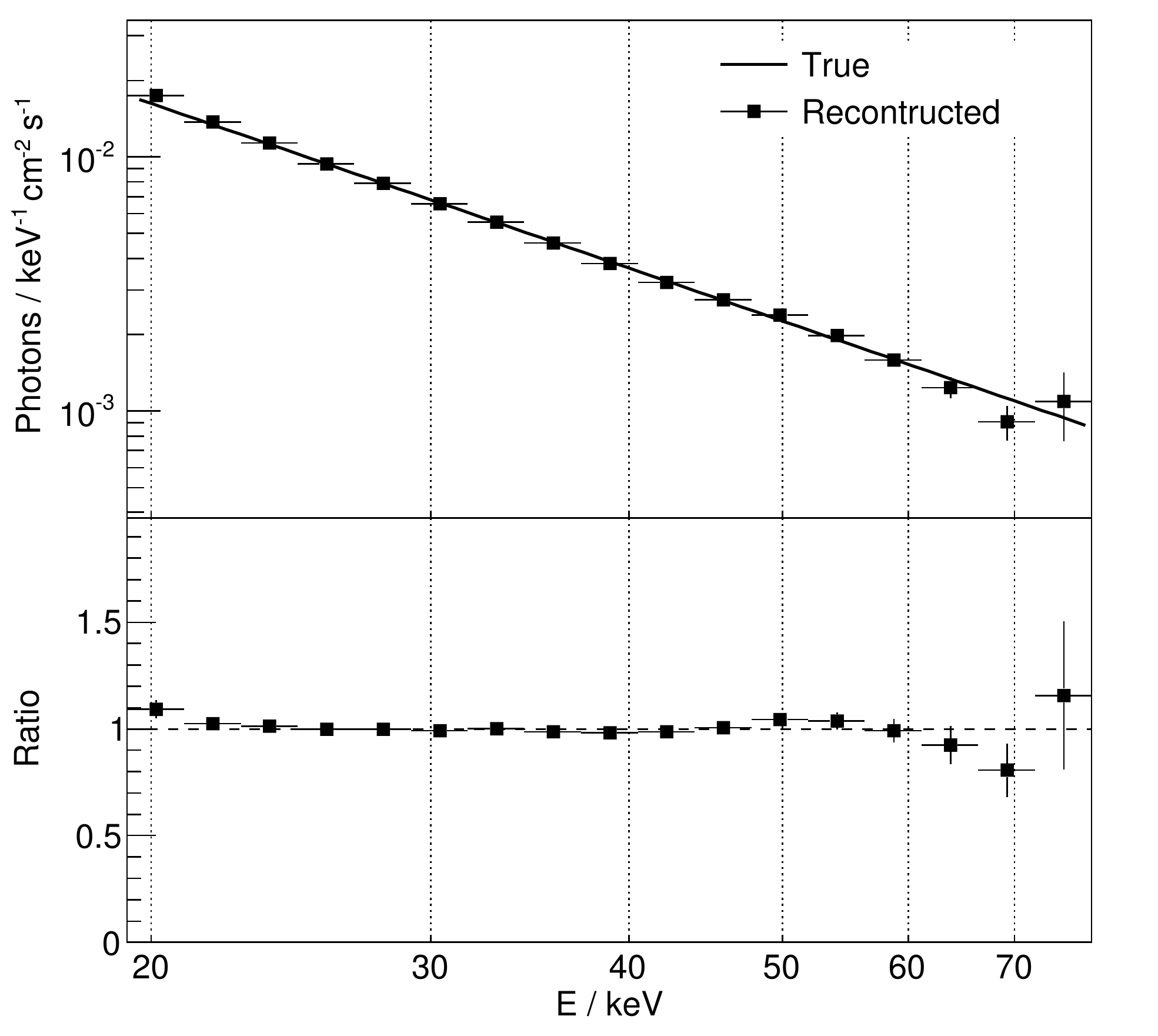}%
    \quad%
    \includegraphics[width=0.45\textwidth]{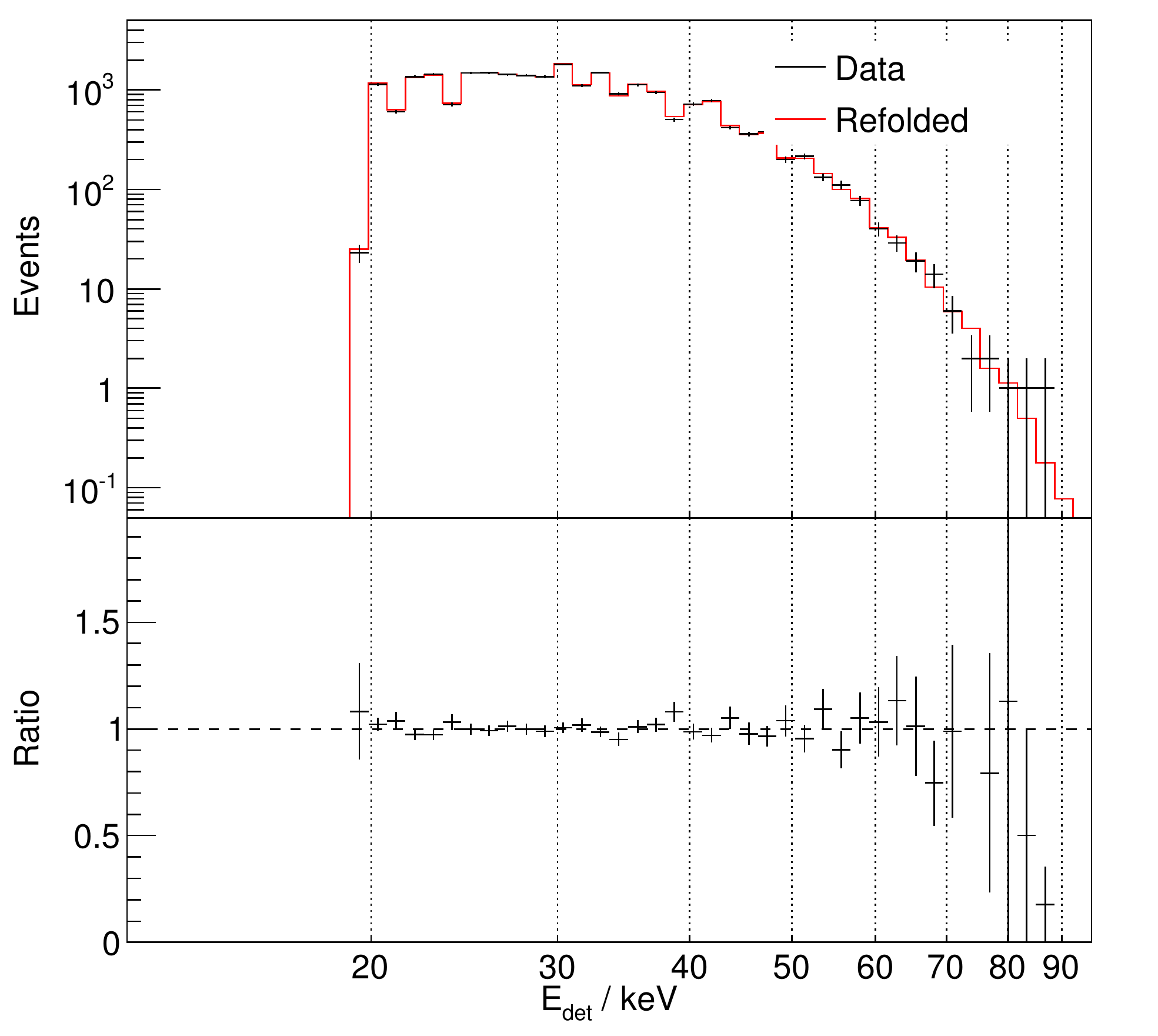}%
    \label{fig:unfolding_result_E-2}%
    \label{fig:refolded_result_E-2}}\\
  \subfloat[Power-law with absorption line starting from~$E^{-2}$ prior.]{
    \includegraphics[width=0.45\textwidth]{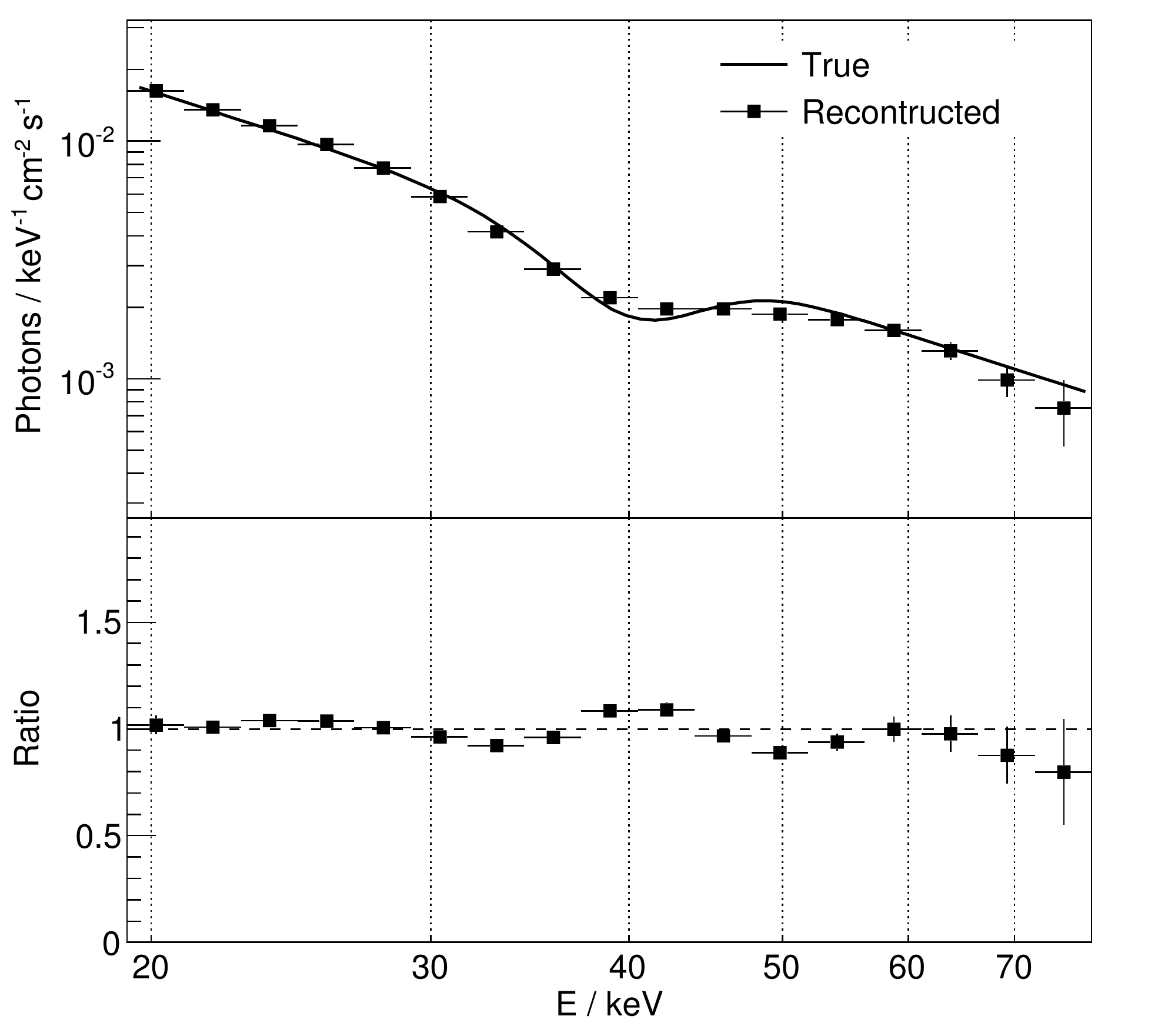}%
    \quad%
    \includegraphics[width=0.45\textwidth]{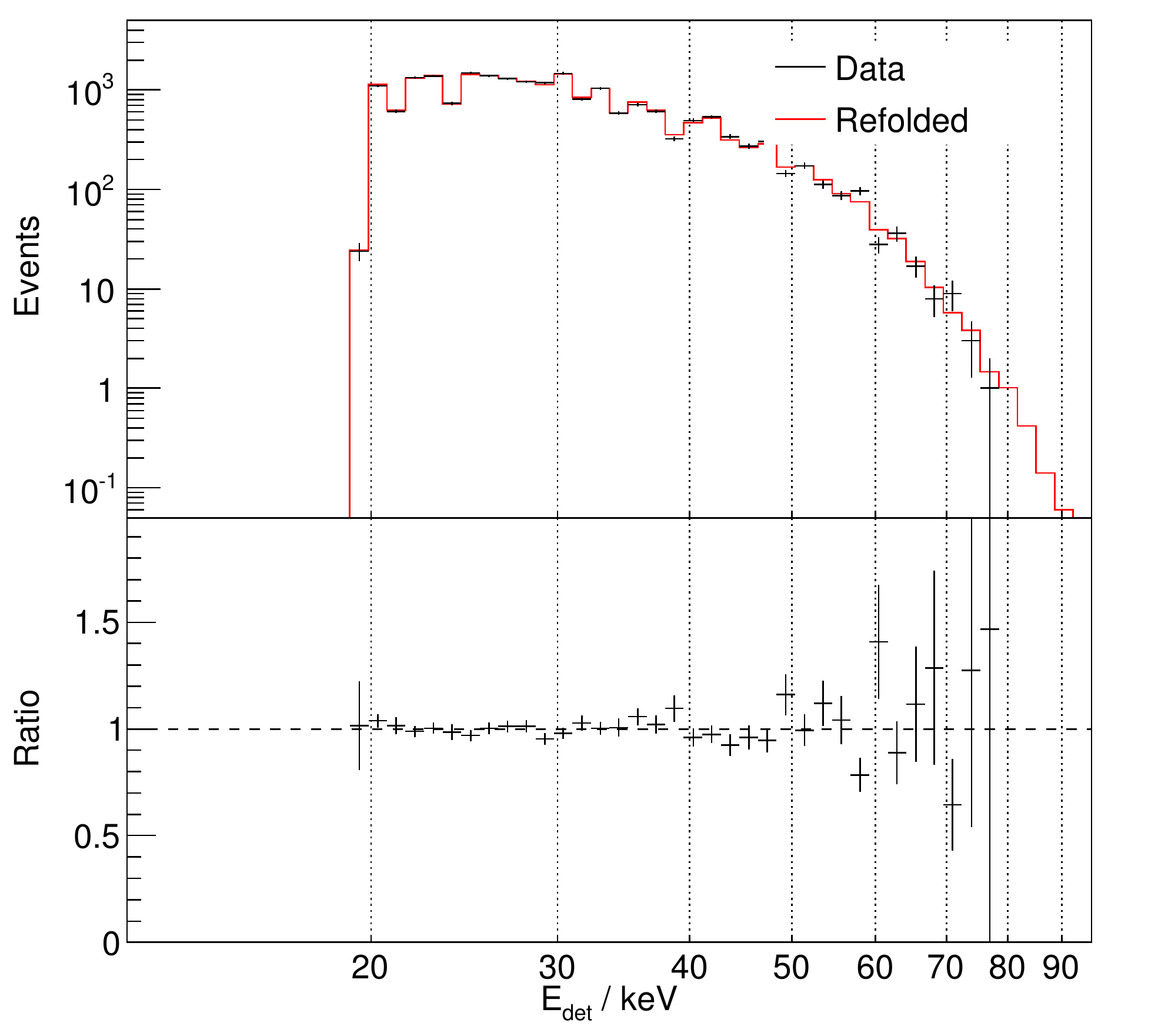}%
    \label{fig:unfolding_result_E-2+line}%
    \label{fig:refolded_result_E-2+line}}
  \caption{Result of the unfolding prodecure starting from an~$E^{-2}$ prior spectrum for (a)~the case of a pure power-law spectrum and (b)~a power-law with an absorption line at an energy of~$40\un{keV}$. Figures~(c) and~(d) show the same but starting from an~$E^{-1}$ prior spectrum. \emph{Left:} Unfolded energy spectrum compared with the true spectrum (see Eq.~\eqref{eq:truespectrum} for the power-law component). \emph{Right:} As a cross-check of the unfolding procedure, the results were folded with the response matrix and compared to the input data. The figures show a projection of the data onto the axis of measured energy.}
  \label{fig:unfolding_result}
  \label{fig:refolded_result}
\end{figure*}

\begin{figure*}[p]
  \ContinuedFloat
  \centering
  \subfloat[Power-law starting from~$E^{-1}$ prior.]{
    \includegraphics[width=0.45\textwidth]{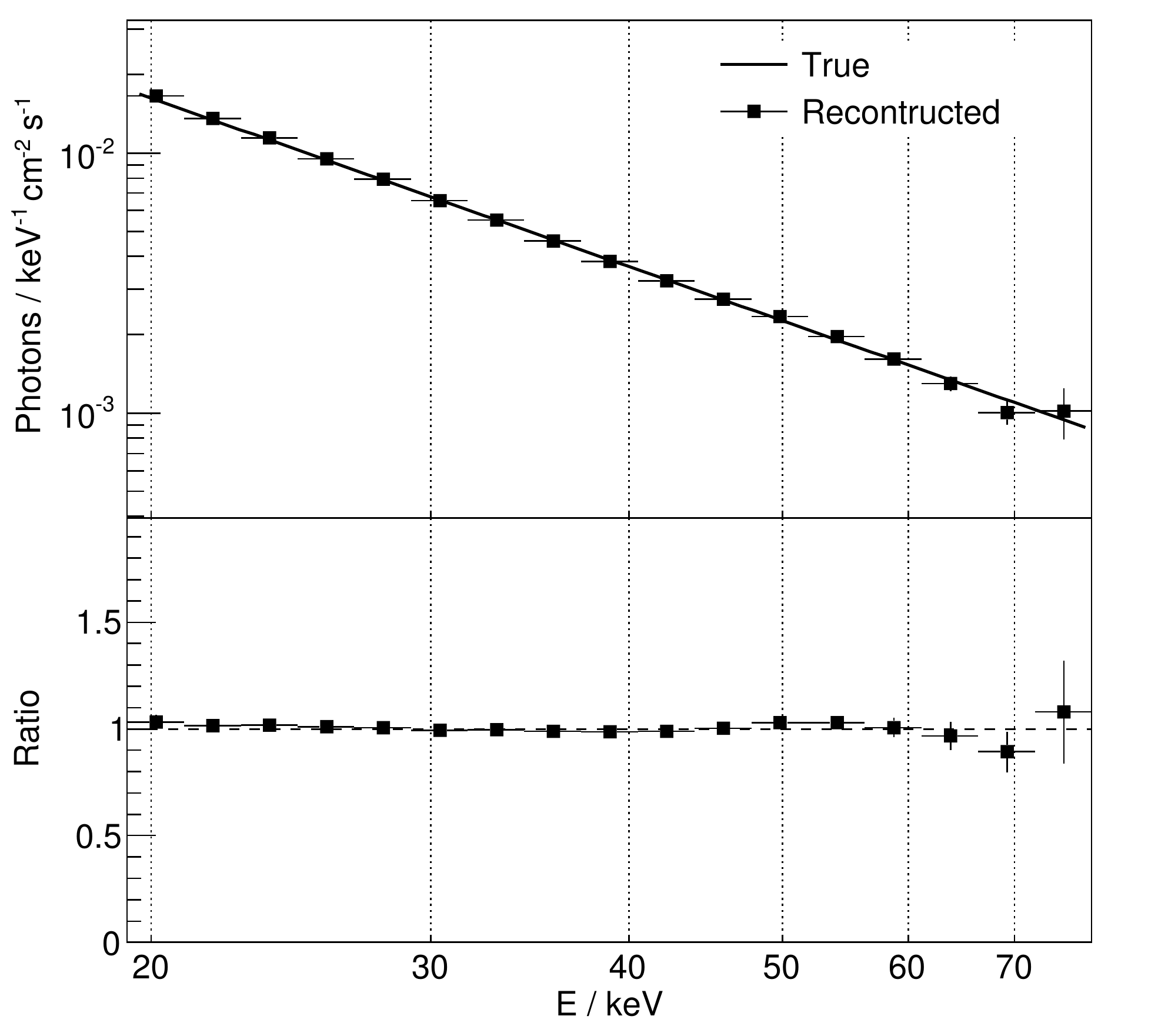}%
    \quad%
    \includegraphics[width=0.45\textwidth]{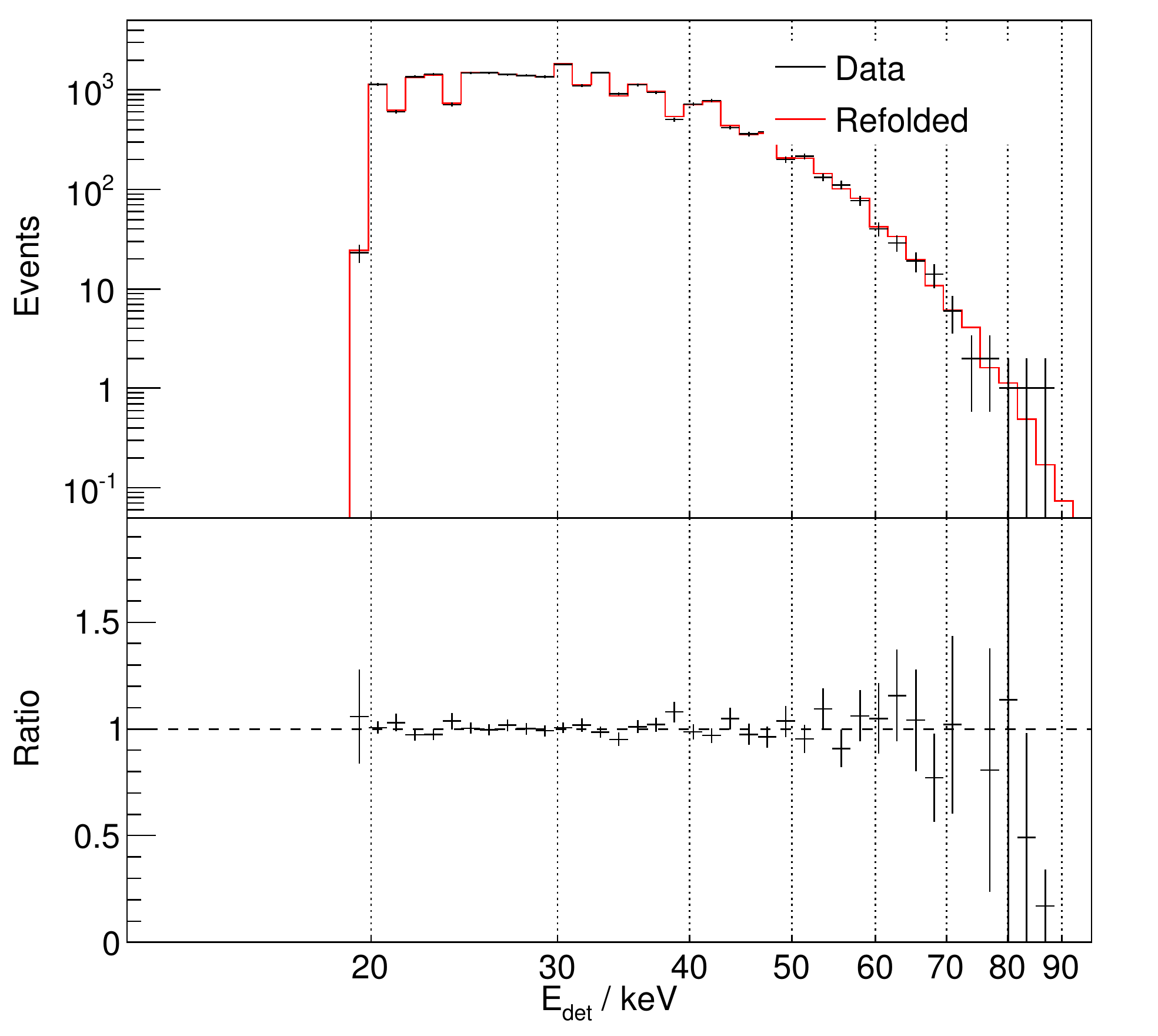}%
    \label{fig:unfolding_result_E-1}%
    \label{fig:refolded_result_E-1}}\\
  \subfloat[Power-law with absorption line starting from~$E^{-1}$ prior.]{
    \includegraphics[width=0.45\textwidth]{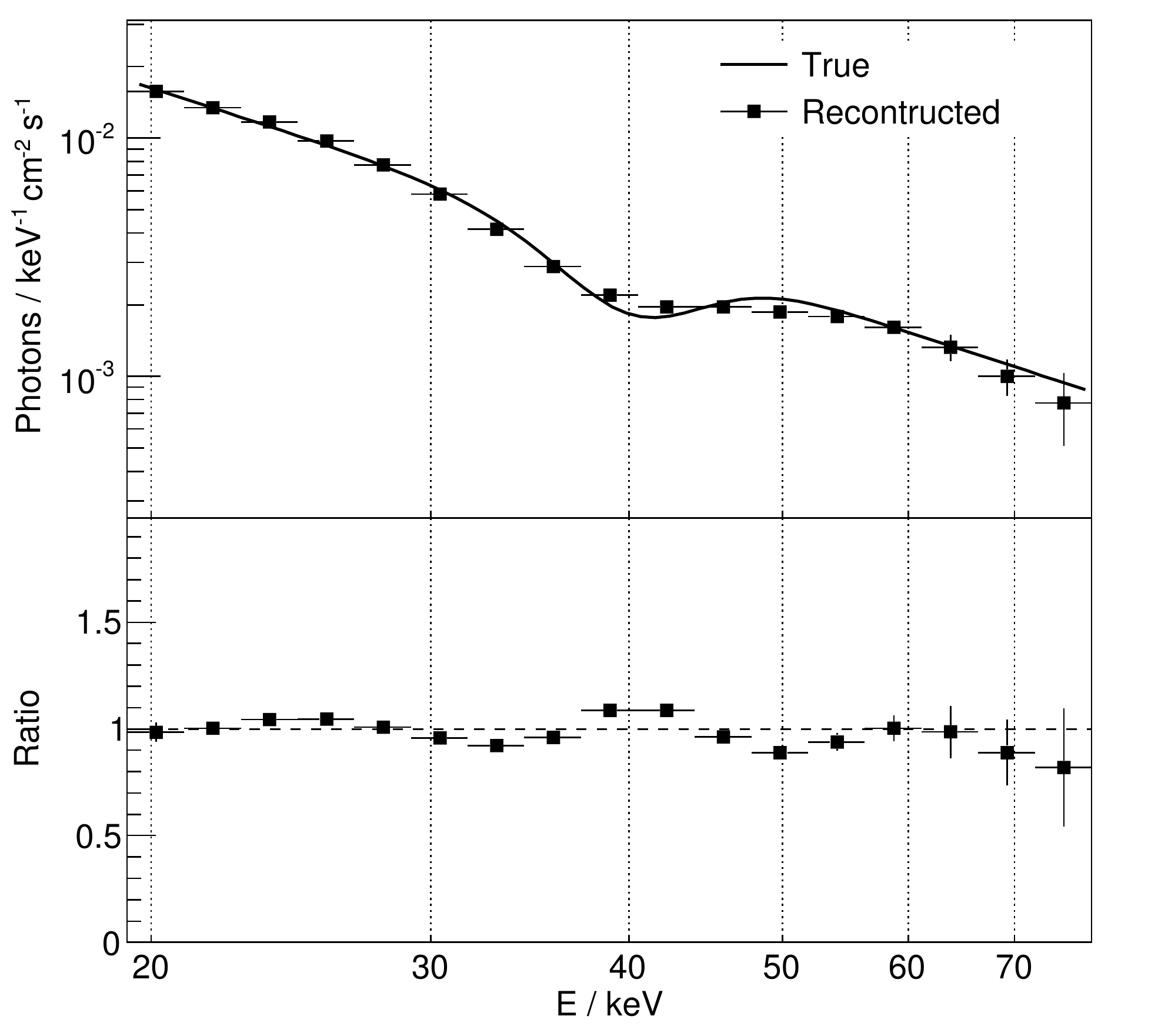}%
    \quad%
    \includegraphics[width=0.45\textwidth]{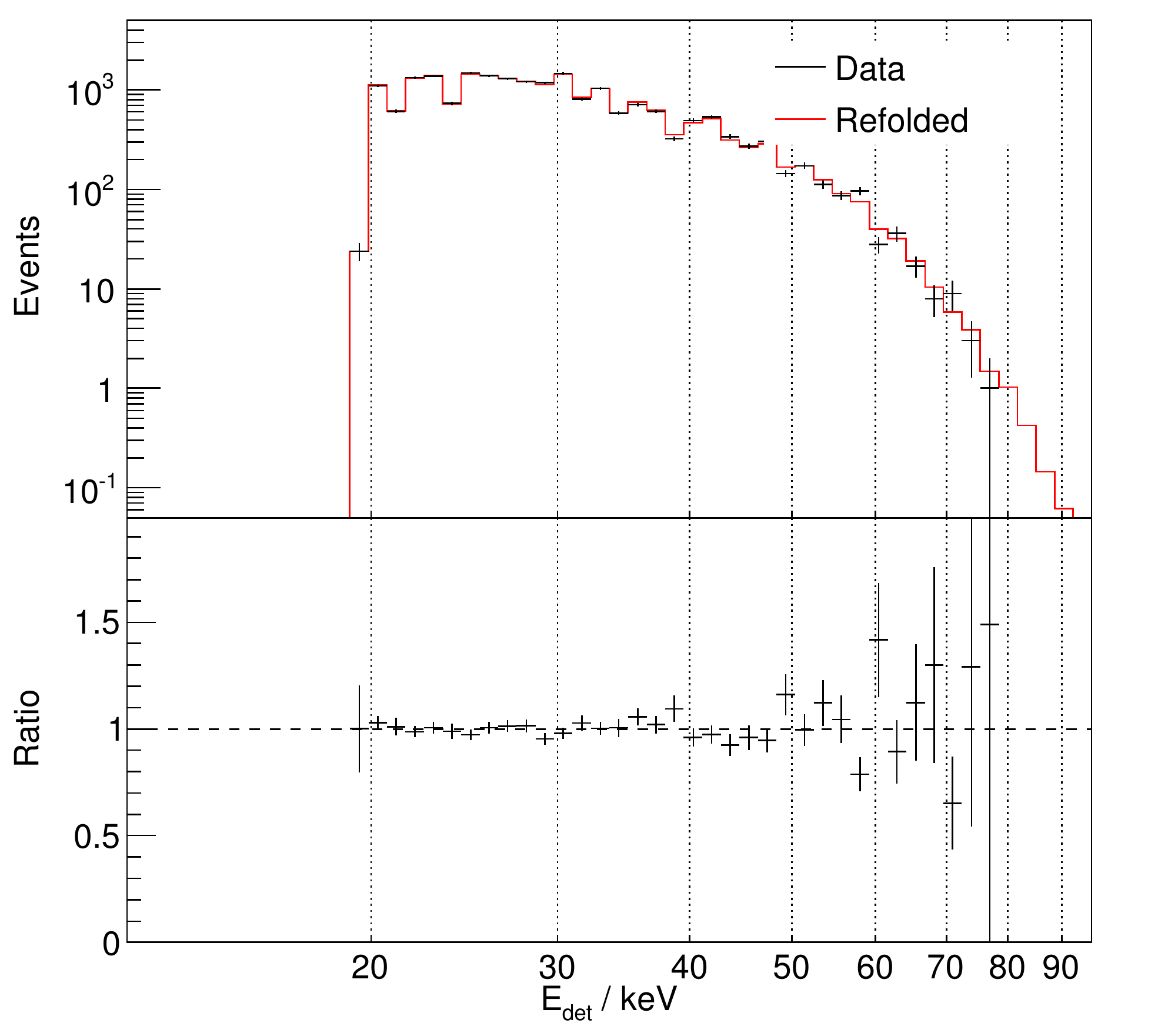}%
    \label{fig:unfolding_result_E-1+line}%
    \label{fig:refolded_result_E-1+line}}
  \caption{See caption on page~\pageref{fig:unfolding_result} for details.}
\end{figure*}

Generally, the energy spectrum is reproduced reasonably well.
The lowest and highest energy bins are problematic due to the low detection efficiencies (see the plots in the right column of Fig.~\ref{fig:unfolding_result}).
The absorption line at~$40\un{keV}$ is smeared out visibly, as seen in Figures~\ref{fig:unfolding_result}\subref{fig:unfolding_result_E-2+line} and~\subref{fig:unfolding_result_E-1+line}.
However, one should note that the width of the line is very close to the energy resolution of the simulated CZT detectors and there is additional energy smearing due to energy losses in the Compton scattering, which cannot be measured in the experiment.
It is important to note that in both cases (pure power-law and power-law plus absorption line) the choice of prior does not have a significant influence on the result beyond the statistical fluctuations.

\begin{figure}
  \centering
  \includegraphics[width=\thisfigurewidth{.5}{.8}]{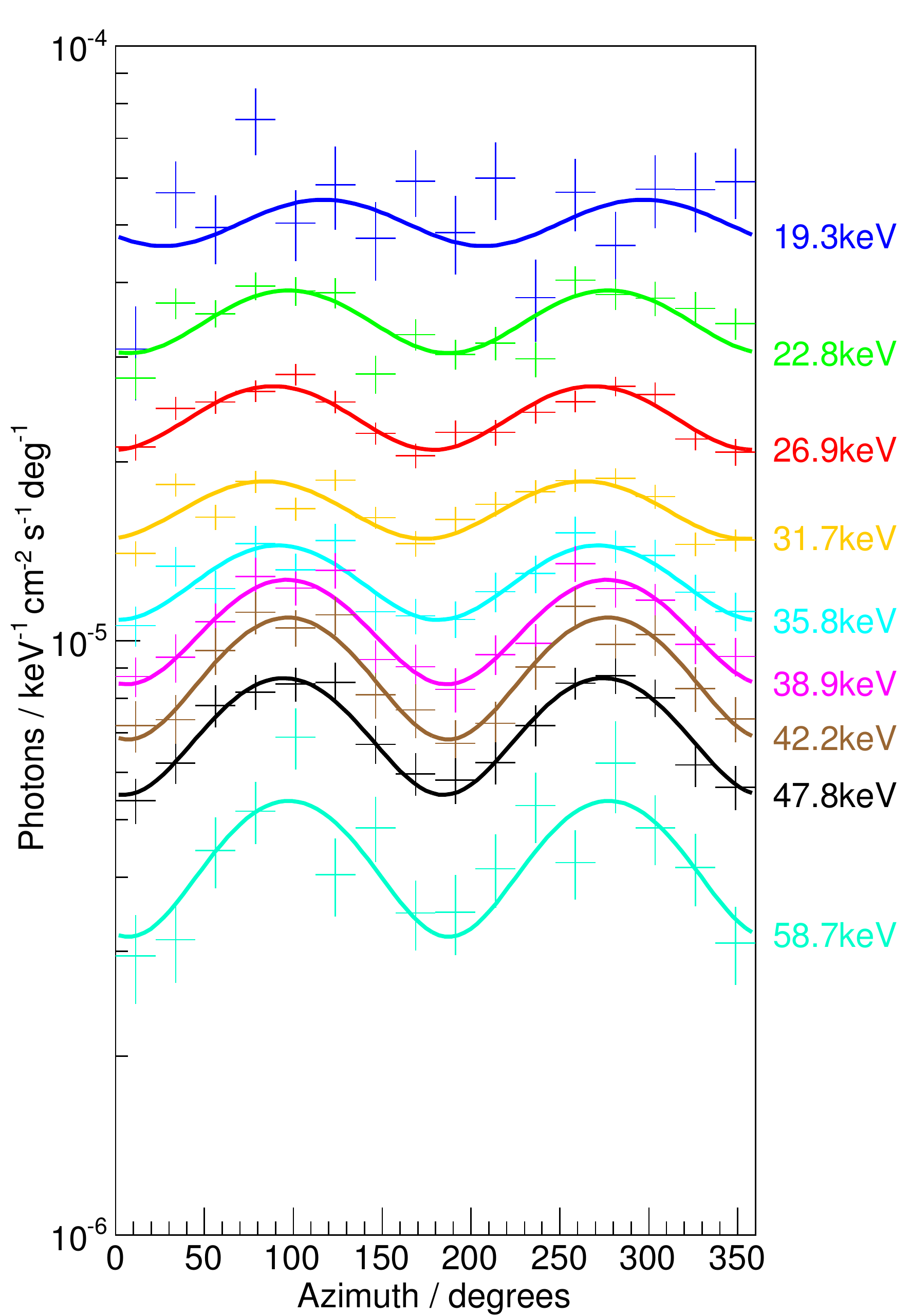}
  \caption{Reconstructed azimuthal distributions in eight energy bins with sine fits for the case of a power-law with absorption line spectrum. The unfolding was started from an~$E^{-2}$ prior. The numbers on the right indicate the central energy of the energy bins.}
  \label{fig:azimuth_fits}
\end{figure}

In a separate step after the unfolding the azimuthal distribution in each energy bin was fitted with a sine function, Eq.~\eqref{eq:sine}, in order to determine the polarization fraction and angle as a function of energy.
In order to improve the accuracy of the determination of polarization parameters, energy bins were combined into larger bins.
The error bars were determined from the uncertainties of the fit parameters through standard error propagation.
As an example, the azimuthal distributions and fits for the case of a power-law with absorption line, starting from an~$E^{-2}$ prior are shown in Fig.~\ref{fig:azimuth_fits}.

\begin{figure}
  \centering
  \includegraphics[width=\figurewidth]{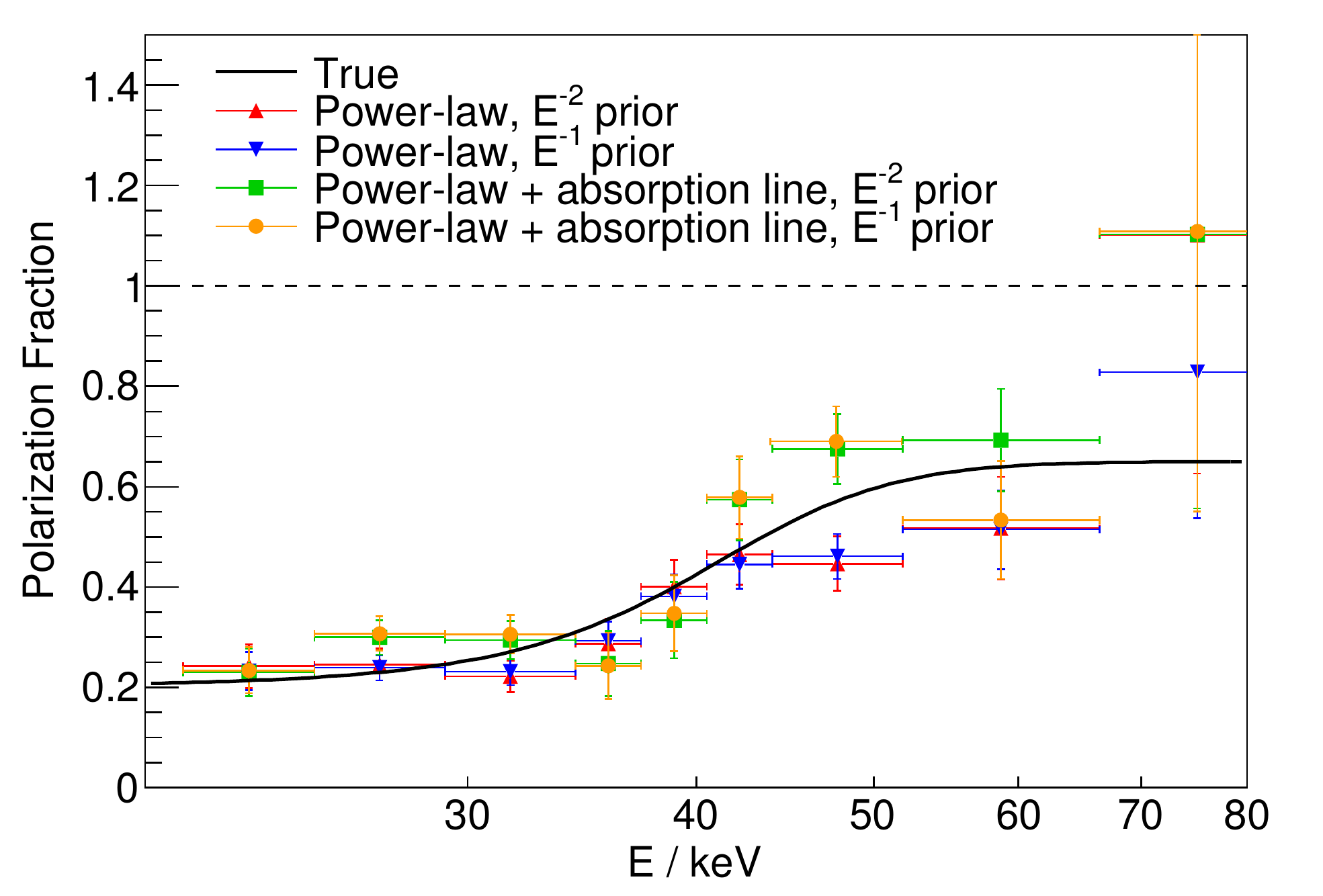}
  \caption{Reconstructed polarization fraction as a function of primary energy after the unfolding prodecure starting from an~$E^{-2}$ and an~$E^{-1}$ prior spectrum model polarization fraction, Eq.~\eqref{eq:truepolarization}.}
  \label{fig:polarization_result}
\end{figure}

The resulting polarization fraction as a function of energy is shown in Fig.~\ref{fig:polarization_result} for all four cases (pure power-law spectrum, power-law plus absorption line, each with an~$E^{-2}$ and a~$E^{-1}$ prior).
In most cases, the change in polarization fraction would be observed by a measurement such as that simulated for this study.
At the highest energies the statistical uncertainties become rather large due to the low statistics in each energy bin.
The results in the individual energy bins are somewhat positively correlated due to the mixing of events in neighboring bins through the unfolding, which leads to slightly larger error bars than one would expect from the scattering of points.
Furthermore, it should be noted that a $\chi^2$ test reveals a slightly better agreement of the results with the model in case of the $E^{-2}$ prior ($\chi^2/N_{df} = 11.9/9$ versus $13.5/9$ and $11.8/9$ versus $14.5/9$).

\section{Summary} \label{sec:summary}
Experiments like X-Calibur, PoGOLite, GEMS, and XIPE are able to measure polarization of X-rays from astrophysical sources over a wide range of energies and enable spectropolarimetric measurements.
Typically, resolution, effective area, and energy response of X-ray polarimeters depend strongly on the energy of the incident X-ray.
Due to the generally steep source spectra, these energy dependencies need to be taken into account when reconstructing the flux, polarization fraction and polarization direction energy spectra.

There are several methods that can be used to achieve this.
Generally, the distributions of input and output variables are related through a response matrix, which describes the probability distribution of outputs for each set of input parameters.
Forward folding is a technique commonly used if a model of the true spectrum should be fitted to the data.
The model spectrum is folded with the response matrix and compared to the measured data, and model parameters are adjusted to achieve a best fit.
The disadvantage of this method is that it cannot deliver model-independent data points.
By combining forward folding with Markov Chain Monte Carlo, the contents of each bin can be treated as an independent parameter allowing a model-independent spectral reconstruction.
However, this becomes very computationally expensive if the number of bins is large, e.\,g.\ in case of multi-dimensional problems.
The inverse of the response matrix can be used to transform the measured distributions into estimates of the true spectrum.
However, direct matrix inversion usually leads to an amplification of statistical fluctuations and consequently a wildly fluctuating result where neighboring values have a strong negative correlation.

In this paper an unfolding-based method is described that makes it possible to reconstruct the flux energy spectrum and to infer the polarization fraction and polarization direction as function of the true energy of the detected photons.
The method makes use of a Bayesian multi-dimensional iterative unfolding procedure~\cite{dagostini_1995}.
For a scattering or photo-effect polarimeter, the input parameters typically are the distributions of measured photon energy and azimuthal scattering angle or azimuth angle of the emitted photo-electron.
The output usually is a two-dimensional distribution of true photon energies and scattering angles.
However, additional parameters can easily be included both on the input and the unfolded data.
In Section~\ref{sec:example}, this is demonstrated for an additional input parameter.

Starting from a prior distribution, which can reflect the best knowledge about or a best guess of the true spectrum and polarization, the data are then unfolded iteratively as described in Section~\ref{sec:spectropolarimetry}.
Statistical uncertainties and the covariance matrix of the unfolding result are computed by varying the input within its statistical errors, and observing variations and correlations of the output spectrum.
In Section~\ref{sec:example} this method was applied to a simulated dataset based on the X-Calibur scattering polarimeter.
Two scenarios were studied: first, the case of a simple power-law spectrum; secondly, a Gaussian shaped absorption line was added at an energy of~$40\un{keV}$.
In both cases the true spectrum was reconstructed reasonably well within 6 or less iterations, in particular when considering that the width of the absorption line is close to the resolution of the simulated detector.
After the unfolding the azimuthal scattering angle distributions were fitted with sine functions in bins of ``true'' energy yielding the polarization fraction and angle as a function of energy.
It was shown that the result does not depend strongly on the chosen prior spectrum, meaning the choice of prior does not lead to a strong bias of the result.

\section*{Acknowledgements}
The authors are grateful for NASA funding from grants NNX10AJ56G \& NNX12AD51G as well as discretionary funding from the McDonnell Center for the Space Sciences.

\section*{References}
\bibliography{xray}
\bibliographystyle{elsarticle-num}

\end{document}